\documentstyle[12pt,amsmath,amsfonts]{article}
\def\note#1{}
\def\ij{\sum_{\stackrel{i,j=1}{\scriptscriptstyle i\neq j}}^N}

\begin{document}
\baselineskip 21pt
\begin{center}

{\Large\bf On the equivalence of the rational\vspace{10pt}\\ 
Calogero--Moser system to free 
particles}

\vskip 1cm

{\large {\bf Tomasz Brzezi\'nski \footnote{Lloyd's Tercentenary Fellow. On
leave from: Department of Theoretical Physics, University of 
\L\'od\'z, ul. Pomorska 149/153, 90-236 {\L}\'od\'z,
Poland. E-mail: tb10@york.ac.uk}}}

\vskip 0.1 cm
Department of Mathematics, University of York\\
Heslington, York YO10 5DD, England\\

\vskip 1cm

{\large {\bf Cezary Gonera} and {\bf Pawe{\l} Ma\'slanka}}

\vskip 0.1 cm

Department of Theoretical Physics II, University of {\L}\'od\'z,\\
ul. Pomorska 149/153, 90-236 {\L}\'od\'z, Poland\\

\end{center}
\vspace{1 cm}
\begin{abstract}
The canonical transformation and its unitary counterpart 
 which relate the rational Calogero--Moser system to the
free one are
constructed. 
\end{abstract}
\thispagestyle{empty}
\newpage

\section{Introduction} \label{sec1}
The Calogero model and its variations \cite{cal69} \cite{sut71} 
\cite{mos75} \cite{ols81} attract recently much attention. They
appear to be of some relevance in various areas of theoretical physics
like quantum Hall effect \cite{kaw93}, fractional statistics \cite{pol89},
two-dimensional gravity \cite{and83} and QCD \cite{min93}, soliton
theory \cite{air77} and the Seiberg--Witten theory \cite{dho97}.

A variety of advanced theoretical tools has been used to gain a deeper
understanding of the structure of Calogero--type models, including
the inverse scattering \cite{ols81} \cite{uji92} and the $r$-matrix methods
\cite{bab90}, $W$-algebra techniques \cite{hik93}, Dunkl's operators
\cite{bri92} and others. 

The rational Calogero--Moser model  is the simplest example of 
this class of
solvable systems.  Its scattering properties
resemble those of a free particle system \cite{ols81} \cite{pol89}. 
On the classical level, the
final configuration of particle momenta coincides with the initial one, 
up to a permutation, and there is no position shift due
to the scattering, while on the quantum level the scattering phase does
not depend on energy. When coupled to the external harmonic potential 
the rational
Calogero--Moser model exhibits the same (up to a common shift) energy spectrum and
degeneracy as the collection of noninteracting harmonic oscillators
\cite{cal69}. The latter property has been explained recently
\cite{gur97} by explicit construction of relevant similarity
transformations relating both systems. For the pure rational Calogero--Moser model
the equivalence to the free one was indicated and shown, in an indirect
way, both for the classical and quantum case, by Polychronakos
\cite{pol89}. However the explicit form of the relevant similarity
transformations cannot be obtained by taking the $\omega =0$ limit of
the transformation in \cite{gur97}, as the latter seems to have an
essential singularity at this point.

In the present note we construct a canonical transformation 
which maps the Hamiltonian of 
the classical Calogero--Moser system to the Hamiltonian of free
particles. We then proceed  to quantise this
construction and thus obtain a unitary
transformation relating the quantum Calogero--Moser and 
free particle systems.
 The construction is based on the
use of the $sl(2,\mathbb{R})$ symmetry characteristic for the  rational
Calogero--Moser models \cite{bar77}. The presented transformation not only relates both
Hamiltonians but it also transforms  the set of symmetric dynamical
variables (the analogs of symmetrised action-angle variables) into
their free counterparts. Furthermore it provides a simple proof of 
the commutation relations conjectured in \cite{pol89}.

\section{The classical case} \label{sec2}
Consider the $N$-particle rational Calogero--Moser model described
by the Hamiltonian 
\begin{equation}
  H_{CM}= \sum_{i=1}^N \frac{p_{i}^{2}}{2}+\frac{g}{2} \sum_{i \neq j} 
  \frac{1}{(q_{i}-q_{j})^2},
  \label{1}
\end{equation}
where $p_i, q_i$ are the canonical variables of the phase space
$\Gamma_N$, and $g$ is a coupling
constant. It is well known \cite{bar77} \cite{gon} that many 
properties of the
Calogero--Moser system can be understood in terms of the representation 
theory of the
Lie algebra 
$sl(2,\mathbb{R})$. Define the following three
functions on the phase space $\Gamma_N$
\begin{subequations}
\begin{eqnarray}
  &&T_+\equiv\frac{1}{\omega} \left(\sum_{i=1}^N \frac{p_{i}^{2}}{2}+
  \frac{g}{2}\sum_{\stackrel{i,j=1}{\scriptscriptstyle i\neq j}}^N\frac{1}{(q_i-q_j)^2}
  \right)=\frac{1}{\omega} H_{CM}, \label{2a}\\
  &&T_-\equiv\omega\sum_{i=1}^N\frac{q_{i}^{2}}{2}, \label{2b}\\
  &&T_0\equiv\frac{1}{2}\sum_{i=1}^Nq_i p_i, \label{2c}
\end{eqnarray}
\end{subequations}
where $\omega$ is a fixed,  nonzero but otherwise arbitrary, frequency.
Using the canonical Poisson brackets on $\Gamma_N$,
$\{q_i,p_j\}=\delta_{ij}$, it is easy to see that 
\begin{equation}
  \{T_0, T_\pm\}=\pm T_\pm, \qquad \{T_+, T_-\}=-2T_0. \label{3}
\end{equation} 
Therefore $T_0, T_\pm$ generate the $sl(2,\mathbb{R})$ (Poisson) algebra. 
\note{Let
\begin{equation}
T_1\equiv\frac{i}{2}(T_+ + T_-), \quad
T_2\equiv\frac{1}{2}(T_+ - T_-), \quad
T_3\equiv T_0, \label{4}
\end{equation}
then (\ref{3}) can be rewritten in the standard $SU(2)$ form
\begin{equation}
  \{T_i, T_j\}=i \varepsilon_{ijk}T_k \label{5}
\end{equation}}
Note that since the Poisson brackets (\ref{3}) do not 
depend on the
coupling constant $g$, also the functions
\begin{equation}
\widetilde{T}_+\equiv\frac{1}{\omega}
    \sum_{i=1}^N\frac{p_{i}^{2}}{2}, \quad
 \widetilde{T}_-\equiv T_-, \quad
 \widetilde{T}_0\equiv T_0 \label{6}
\end{equation}
span the $sl(2,\mathbb{R})$ algebra. 

Next, consider the following one-parameter family of
transformations 
\begin{eqnarray}
  && q_k\rightarrow \mbox{e}^{i\lambda T_1}\ast q_k\equiv 
     \sum_{n=0}^{\infty} \frac{(i\lambda)^n}{n!} 
     \left\{T_1,\dots\left\{T_1, q_k\right\}\dots\right\},\nonumber\\
  && p_k\rightarrow \mbox{e}^{i\lambda T_1}\ast p_k\equiv 
     \sum_{n=0}^{\infty} \frac{(i\lambda)^n}{n!} 
     \left\{T_1,\dots\left\{T_1, p_k\right\}\dots\right\}, \label{7}
\end{eqnarray}
where $T_1 = \frac{i}{2}(T_+ + T_-)$. These are well-defined, real 
canonical transformations on 
$\Gamma_N/S_q$, where $S_q\subset\Gamma_N$ consists of points $(q,p)$ such
that $q_i=q_j$ for at least one pair of indices $i\neq j$. In fact,
eqs.\ (\ref{7}) simply define the time evolution $(t\equiv
\frac{\lambda}{2\omega})$ generated by the Hamiltonian of
the Calogero--Moser model coupled to the external harmonic force
characterized by the frequency $\omega$
\begin{equation}
  H_C\equiv -\frac{2i}{\omega}T_1=\sum_{i=1}^N\left(\frac{p_{i}^{2}}{2}+
  \frac{\omega^2 q_{i}^{2}}{2}\right)+\frac{g}{2}\ij
  \frac{1}{(q_i-q_j)^2}. \label{8}
\end{equation}
In the following the model corresponding to the Hamiltonian $H_C$ in
(\ref{8})  will be called {\em the Calogero model}
 (it differs
trivially from the original Calogero model \cite{cal69}).

On the other hand it follows immediately from the Poisson brackets
(\ref{3}) that the canonical  transformation (\ref{7}) 
rotates the space spanned by $T_0, T_\pm$
around the 
axis $T_1$ by an angle $\lambda$. Therefore we
obtain 
\begin{equation}
  \omega T_+(q,p)\rightarrow (\mbox{e}^{i\pi T_1}\ast T_+)(q,p)=
  \omega T_-(q,p)=\omega \widetilde{T}_-(q,p). \label{9}
\end{equation}
Now, one can use the canonical transformation generated by
$\widetilde{T}_1= \frac{i}{2}(\widetilde{T}_+ + \widetilde{T}_-)$ 
to rotate $\widetilde{T}_-$ back to
$\widetilde{T}_+$
\begin{equation}
  \omega \widetilde{T}_-(q,p)\rightarrow 
  \left(\mbox{e}^{-i\pi \widetilde{T}_1}\ast
  \widetilde{T}_-\right)(q,p)=
  \omega \widetilde{T}_+(q,p)=\sum_{i=1}^N\frac{p_{i}^{2}}{2}. \label{10}
\end{equation}
This is again a well-defined canonical transformation which is a time
transformation by $t=-\frac{\pi}{2\omega}$ generated by the harmonic
oscillators Hamiltonian. It can be described explicitly by the
following formulae:
\begin{equation}
q_k\rightarrow -\frac{1}{\omega} p_k, \qquad
 p_k\rightarrow \omega q_k. \label{11}
\end{equation}
Thus we conclude that the canonical transformation 
\begin{equation}
 q_k\rightarrow \mbox{e}^{-i\pi \widetilde{T}_1}\ast
     \left(\mbox{e}^{i\pi T_1}\ast q_k\right), \qquad
  p_k\rightarrow \mbox{e}^{-i\pi \widetilde{T}_1}\ast
     \left(\mbox{e}^{i\pi T_1}\ast p_k\right) \label{12}
\end{equation}
transforms the Calogero--Moser model into the free particle theory.
Since this
transformation does not depend explicitly on time, both
systems are equivalent to each other.

When applied to the Hamiltonian (\ref{1}), the transformation (\ref{12})
  results
effectively in setting $g=0$. We devote the rest of this section to
showing  that this formal statement
remains true for a
rather wide class of observables. It is well known \cite{ols81} 
\cite{pol89} that if one  restricts oneself to the observables which are
symmetric functions on the phase space, the dynamics of the Calogero--Moser
system is completely described by the following set of dynamical
variables. Begin with the integral of motion in the Henon form
\cite{saw75} \cite{woj77} 
\begin{equation}
  I_N\equiv \mbox{e}^{-\frac{g}{2} \sum\limits_{i\neq j}
  \frac{1}{(q_i-q_j)^2} \frac{\partial^2}{\partial p_i \partial p_j}}
  \prod_{k=1}^N p_k. \label{13}
\end{equation}
The full set of mutually commuting independent integrals of motion
can be then obtained by taking the successive Poisson brackets of $I_N$
with $\sum\limits_{i}q_i$ \cite{woj77}, i.e.,  
\begin{equation}
  I_{N-n} \equiv \frac{1}{n!} 
  \biggl\{
  \underbrace{\sum_{i=1}^N q_i, \dots
  \biggl\{
  \sum_{i=1}^N q_i}_{\mbox{\small n times}}, I_N 
  \biggr\}
  \dots
  \biggr\}, 
  \quad n=0,1,\dots,N-1. \label{14} 
\end{equation}
More generally, one defines \cite{pol89} \cite{bar77} \cite{gon}
\begin{eqnarray}
  I_{m\, n}\equiv \frac{1}{2^m m!} \frac{1}{(N-m-n)!} 
  \biggl\{
  \underbrace{\sum_{i=1}^N q_{i}^{2}, \dots 
  \biggl\{
  \sum_{i=1}^N q_{i}^{2}}_{\mbox{\small m times}},
  \biggl\{
  \underbrace{\sum_{i=1}^N q_i, \dots 
  \biggl\{
  \sum_{i=1}^N q_i}_{\mbox{\small N-m-n times}}, I_N
  \biggr\}
  \biggr.
  \biggr.
  \dots
  \biggr\},  && \label{15}\\
  && \nonumber\\
  1\leq m+n \leq N. && \nonumber 
\end{eqnarray}
The functions $I_{mn}$ (\ref{15}) obey \cite{pol89}
\begin{equation}
  \dot{I}_{m\, n}=m I_{m-1\, n+1}. \label{16}
\end{equation}
This implies that all the $I_{mn}$ have a polynomial time dependence. 
Actually, not all functions $I_{mn}$ are needed for the description of
the dynamics of the system. Modulo particle permutations, 
the system is fully described by
the following $2N$ quantities
\begin{eqnarray}
  I_n\equiv I_{0\,n},\qquad
  J_n\equiv I_{1\,n-1}=\frac{1}{2}
  \biggl\{\sum_{i=1}^N q_{i}^{2}, J_n \biggr\}. \label{17}
\end{eqnarray}
By  eqs.\ (\ref{16}), the functions $I_n$ are constants of motion,
 while the functions $J_n$
depend linearly on time. The functions $J_n$  
can be used to construct new $N-1$
independent integrals of motion which do not depend explicitly on
time, thus showing the maximal superintegrability of the 
rational Calogero--Moser
systems \cite{woj83} \cite{gon?}. Moreover, again up to
the particle permutation, they determine
 the asymptotic form of the particle motion,
i.e. they fix the trajectory.

We now show that the canonical transformation (\ref{12}) 
applied to the quantities (\ref{17}) amounts to putting $g=0$,
i.e. the transformed observables read
\begin{equation}
  I_n(g=0)=\sum_{\stackrel{i_1,\ldots , i_n = 1}{\scriptscriptstyle 
i_1<\dots <i_n}}^N \prod_{k=1}^{n} p_{i_k}, \qquad
  J_n(g=0)=\sum_{j=1}^N q_j 
  \biggl( 
  \sum_{\substack{i_1<\dots<i_{n-1}\\
                  i_1\neq j\dots i_{n-1}\neq j}}
  \prod_{k=1}^{n-1} p_{i_k}
  \biggr). \label{18}
\end{equation}
It should be, however, stressed that, for the reasons which will be
explained below, this conclusion does not apply to
$I_{m\,n}$ for $m\geq 2$. 

The first observation to make is that once it is proven that the
transformed $I_N$ has the form stated in (\ref{18}), i.e., $I_N \to 
\Pi_{k=1}^Np_k$, the relations for all other $I_n$ will follow. This
can be seen as follows. {}From the fact that the centre-of-mass
motion is insensitive to the internal  forces it is clear that 
$\sum\limits_{k=1}^{N} q_k$ is invariant under
(\ref{12}). The Poisson bracket
is invariant under the  canonical transformations, therefore
\begin{equation}
  I_{N-n} \longrightarrow \frac{1}{n!} 
  \biggl\{\sum_{i=1}^N q_i,\dots \biggl\{\sum_{i=1}^N q_i, \prod_{k=1}^N p_k
  \biggr\}\dots \biggr\}=I_{N-n}(g=0), \label{29}
\end{equation}
provided $I_N$ transforms as stated. 

In order to find the transformation properties
of $I_N$ consider the Calogero model given by the Hamiltonian $H_C$ in
(\ref{8}). It has been shown in \cite{gon} that the
functions $\phi_k(q,p)$, $k=-\frac{N}{2},
-\frac{N}{2},\dots,\frac{N}{2}$ given by 
\begin{equation}
  \phi_k=\frac{1}{N!}\sum_{n=0}^{N} (N-n)! 
  C_{N-n}\biggl(\frac{N}{2},k\biggr) 
  \biggl(
  \omega \sum_{i=1}^{N} q_i \frac{\partial}{\partial p_{i}} 
  \biggr)^n I_N, \label{20}
\end{equation}
where the coefficients $C_n\biggl(\frac{N}{2},k\biggl)$ are defined
via the relation
\begin{equation}
  (x+i)^{\frac{N}{2}+k} (x-i)^{\frac{N}{2}-k}=\sum_{n=0}^{N} 
  C_n\biggl(\frac{N}{2},k\biggr) x^n, \label{21}
\end{equation}
have the following simple time behaviour
\begin{equation}
  \phi_k(t)=\mbox{e}^{2ik\omega t} \phi_k(0). \label{19}
\end{equation}
Since the canonical transformation (\ref{7}) corresponds to the time
shift, the relation (\ref{19}) implies that 
\begin{equation}
  \phi_k\longrightarrow \mbox{e}^{i\pi T_1}\ast
  \phi_k=\mbox{e}^{ik\pi}\phi_k. \label{24}
\end{equation}
The idea of deriving of transformation properties of function $I_N$ 
is to express it in terms of $\phi_k$ using eqs.\ (\ref{20}), and then to 
use the simple form (\ref{24}) 
of the transformation of $\phi_k$.  To implement this idea observe that
the identities
\begin{subequations}
\begin{eqnarray}
  (2x)^N & = &((x+i)+(x-i))^N=
  \sum_{k=-\frac{N}{2}}^{\frac{N}{2}} 
  \binom{N}{\frac{N}{2}+k} (x+i)^{\frac{N}{2}+k} 
  (x-i)^{\frac{N}{2}-k}, \label{22a}\\
  (2i)^N & = &((x+i)-(x-i))^N\nonumber\\
  & = &\sum_{k=\frac{N}{2}}^{\frac{N}{2}} 
  \binom{N}{\frac{N}{2}+k}^{\frac{N}{2}+k} 
  (x+i)^{\frac{N}{2}+k} (x-i)^{\frac{N}{2}-k} 
  (-1)^{\frac{N}{2}-k}, \label{22b}
\end{eqnarray}
\end{subequations}
together with (\ref{21}) imply that
\begin{subequations}
\begin{eqnarray}
  &&\sum_{k=-\frac{N}{2}}^{\frac{N}{2}} \binom{N}{\frac{N}{2}+k}
  C_n\biggl(\frac{N}{2},k \biggr)=2^N \delta_{n\,N}, \label{23a}\\
  &&\sum_{k=-\frac{N}{2}}^{\frac{N}{2}} \binom{N}{\frac{N}{2}+k}
  \mbox{e}^{ik\pi}C_n\biggl(\frac{N}{2},k \biggr)=
  (-2)^N \delta_{n\,0}. \label{23b}
\end{eqnarray}
\end{subequations}
The definition of $\phi_k$  (\ref{20}) together with eqs.\ 
(\ref{23a}), (\ref{23b}) lead to
\begin{equation}
  2^{-N}\sum_{k=-\frac{N}{2}}^{\frac{N}{2}} \binom{N}{\frac{N}{2}+k} 
  \phi_k=I_N \label{26}
\end{equation}
and
\begin{eqnarray}
  &&\hspace{-1cm}2^{-N}\sum_{k=-\frac{N}{2}}^{\frac{N}{2}}
  \binom{N}{\frac{N}{2}+k} {\mbox e}^{ik\pi} \phi_k=\frac{(-1)^N}{N!} 
  \biggl(\omega\sum_{i=1}^{N}q_i 
  \frac{\partial}{\partial p_i}\biggr)^N I_N=\nonumber\\
  &&\hspace{-1cm}\mbox{e}^{-\frac{g}{2}\sum
  \limits_{i\neq j} \frac{1}{(q_i-q_j)^2} 
  \frac{\partial^2}{\partial p_i \partial p_j}}
  \biggl(\frac{(-\omega)^N}{N!}\biggl(\sum_{i=1}^{N}q_i 
  \frac{\partial}{\partial p_i}\biggr)^N \prod_{k=1}^{N} p_k\biggr)=
  (-\omega)^N \prod_{k=1}^{N} q_k \label{27}
\end{eqnarray}
Since a canonical transformation is a linear transformation, combining
(\ref{24}) with (\ref{26}) and (\ref{27}) and then using eqs.\ (\ref{11})
we finally obtain 
\begin{equation}
  I_N \longrightarrow \mbox{e}^{-i\pi \widetilde{T}_1}\ast 
  \biggl(\mbox{e}^{i\pi T_1}\ast I_N\biggr)=
  \prod_{k=1}^{N} p_k, \label{28}
\end{equation}
as stated. This completes the proof of the transformation rules
(\ref{18}) for all the $I_n$. 
On the other hand, due to the relation 
\begin{equation}
  \frac{1}{2} \sum_{i=1}^N q_{i}^{2}=\frac{1}{\omega} T_-, \label{30}
\end{equation}
we have
\begin{equation}
  \frac{1}{2}\sum_{i=1}^N q_{i}^{2} \longrightarrow 
  \biggl(\sum_{i=1}^N\frac{q_{i}^{2}}{2}+\frac{g}{2}\ij 
  \frac{1}{(p_i-p_j)^2}\biggr) \label{31}
\end{equation}
and
\begin{eqnarray}
  \biggl\{\frac{1}{2}\sum_{i=1}^Nq_{i}^{2}, I_n\biggr\} \longrightarrow 
  &&\hspace{-0.6cm}\biggl\{\sum_{i=1}^N\frac{q_{i}^{2}}{2}+\frac{g}{2}
  \ij\frac{1}{(p_i-p_j)^2}, I_n(g=0)\biggr\}=\nonumber\\
  &&\hspace{-0.6cm}=\biggl\{\sum_{i=1}^N\frac{q_{i}^{2}}{2}, 
  I_n(g=0)\biggr\}, \label{32}
\end{eqnarray}
which, together with (\ref{17}), proves the assertion for $J_n$.
Notice that this proof does not apply to $I_{n\,m}$ with $m\geq 2$. In
fact, $\biggl\{\sum\limits_{i} q_{i}^{2},I_n(g=0)\biggr\}$ depends on
the $q_i$ and the term 
$\frac{g}{2}\sum\limits_{i\neq j}\frac{1}{(p_i-p_j)^2}$ cannot be neglected 
 when taking successive
Poisson brackets. In order to understand why this is the case consider
the motion of the Calogero--Moser particles in the limit 
$t\rightarrow -\infty$, say. Due to the repulsive character of the
internal forces the particles are well-separated
and move freely in this limit, i.e., $p_i\simeq p_{i}^{-}$, 
$q_i\simeq p_{i}^{-}t+a_{i}^{-}$. 
Moreover, one can put $g=0$ when calculating
$I_n$ and $J_n$ from the asymptotic data. Therefore we conclude
that the canonical transformation (\ref{12}) transforms the motion of
the Calogero--Moser
particles into free motion which coincides, up to the particle
permutation, with $t\rightarrow -\infty$ asymptotics of the Calogero--Moser
trajectory. Note in passing that the property 
$p_{i}^{-}\neq p_{j}^{-}$, $i\neq j$ implies that 
the image of the transformation
(\ref{12}) is $\Gamma_N/S_p$, where, $S_p$ is defined in the same way 
as $S_q$ with
$q_i$ replaced by $p_i$.

It is now easy to understand why the reasoning concerning the form of
transformed $I_n$ and $J_n$ does not work for $I_{n\,m}$ with 
$m\geq 2$. The trajectory of the $i$-th Calogero--Moser particle can be written as
\begin{equation}
  q_i(t)=p_{i}^{-}t+a_{i}^{-}+O\biggl(\frac{1}{t}\biggr) \label{33}
\end{equation}
where the term $O(\frac{1}{t})$  depends on 
$p_{j}^{-}$, $a_{i}^{-}$ and $g$ in general. In the product
$q_{i}(t)q_{j}(t)$ the term linear in $t$, when multiplied by 
the $O(\frac{1}{t})$ term produces additional nonvanishing
$g$-dependent contribution, absent in the free particles case. 
For $J_n$ this argument is no longer applicable. Although still
we have 
\begin{equation}
  p_{i}(t)=p_{i}^{-}+O\biggl(\frac{1}{t}\biggr) \label{34}
\end{equation}
in the coefficient in front of $t$ all the $O(\frac{1}{t})$ contributions
cancel because this coefficient is an exact integral of motion.

Another way of looking at this phenomenon is to
realise that the (nonlinear) algebraic relations between $I_{n\,m}$, 
$m\geq 2$ and the $I_n$ and $J_n$ contain $g$ explicitly.

\note{\begin{equation}
  2^{-N}\sum_{k=-\frac{N}{2}}^{\frac{N}{2}} \binom{N}{\frac{N}{2}+k}
  \phi_k \longrightarrow 2^{-N}\sum_{k=-\frac{N}{2}}^{\frac{N}{2}}
  \binom{N}{\frac{N}{2}+k} \phi_k\mbox{e}^{ik\pi} \label{25}
\end{equation}}

We conclude this section by computing the canonical transformation
(\ref{12}) and its action on functions $I_{mn}$ in the case of the
2-particle Calogero--Moser model. Separating the
centre-of-mass motion
\begin{eqnarray}
  &&q\equiv q_1-q_2, \quad p\equiv \frac{1}{2}(p_1-p_2) \nonumber\\
  &&X\equiv \frac{1}{2} (q_1+q_2), \quad \Pi\equiv p_1+p_2, \label{35}
\end{eqnarray}
one obtains
\begin{equation}
T_+=\frac{1}{\omega}\biggl(\frac{\Pi^2}{4}+
  \biggl(p^2+\frac{g}{q^2}\biggr)\biggr), \quad
T_0=\frac{1}{2}X\Pi+\frac{1}{2}qp, \quad
 T_-=\omega X^2+\frac{\omega}{4}q^2. \label{36}
\end{equation}
Obviously, $X$ and $\Pi$ are invariant under (\ref{12}). For the
relative coordinates we solve first the equations of motion for the
Hamiltonian $H_C=p^2+\frac{g}{q^2}+\frac{\omega^2}{4}q^2$. We obtain
\begin{subequations}
\begin{eqnarray}
  &&q^2(t)=\frac{2E}{\omega^2}\biggr(1+\frac{\omega}{E}q(0)p(0)\sin
  2\omega t+\biggl(\frac{\omega^2 q^2(0)}{2E}-1\biggr)\cos 2\omega
  t\biggr) \label{37a}\\
  &&q(t)p(t)=\frac{E}{\omega^2}\biggl(\frac{\omega^2}{E}q(0)p(0)\cos
  2\omega t+\biggl(\frac{\omega^2 q^2(0)}{2E}-1\biggr)\sin 2\omega
  t\biggr) \label{37b}\\
  &&E=p^2(0)+\frac{g}{q^2(0)}+\frac{\omega^2}{4}q^2(0). \label{37c} 
\end{eqnarray}
\end{subequations}
Since $q(0)>0$, ($q(0)<0$ resp.) implies $q(t)>0$, ($q(t)<0$ resp.),
the  canonical transformation (\ref{12}) comes out as
\begin{equation}
  q^\prime=\frac{p\,q\,\mbox{sgn}(q)}{\sqrt{p^2+\frac{g}{q^2}}}, \qquad
  p^\prime=\mbox{sgn}(q)\sqrt{p^2+\frac{g}{q^2}}, \label{38}
\end{equation}
and is well-defined for $q\neq 0$. The inverse transformation reads
\begin{equation}
  q=\mbox{sgn}(p^\prime)\sqrt{q^\prime+\frac{g}{{p^\prime}^2}}, \qquad
  p=\frac{q^\prime\,{p^\prime}^2}{\sqrt{g+{q^\prime}^2 
  {p^\prime}^2}} \label{39}
\end{equation}
and is defined for $p^\prime\neq 0$. Both transformations do not
depend on $\omega$.

In order to check that (\ref{38}) and (\ref{39}) transform the Calogero--Moser
system into a free one we write out the explicit solutions of the Calogero--Moser
equations of motion
\begin{eqnarray}
  &&q(t)=\mbox{sgn}(q(0))\sqrt{\frac{g+(2Et+q(0)p(0))^2}{E}}
  \nonumber\\
  &&p(t)=\mbox{sgn}(q(0))
  \frac{(2Et+q(0)p(0))}{\sqrt{\frac{g+(2Et+q(0)p(0))^2}{E}}}
  \label{40}\\
  &&E\equiv p^2(0)+\frac{g}{q^2(0)} \nonumber  
\end{eqnarray}
The asymptotic form for $q(t)$ $(t\rightarrow -\infty)$ reads 
\begin{equation}
  \mbox{\normalsize $
  q(t)=-\mbox{sgn}(q(0))\Bigl(2\sqrt{E}t+\frac{q(0)p(0)}{\sqrt{E}}+
  \frac{g}{2\sqrt{E}(2Et+q(0)p(0))}\Bigr)+0\Bigr(\frac{1}{t^2}\Bigl)$}
  \label{41}
\end{equation}
i.e.
\begin{equation}
  p^-=-\mbox{sgn}(q(0))\sqrt{E}, \qquad 
 a^-=-\mbox{sgn}(q(0))\frac{q(0)p(0)}{\sqrt{E}} \label{42}
\end{equation}
On the other hand, inserting (\ref{40}) into the right-hand side of
 (\ref{38}), one obtains
\begin{equation}
  q^\prime(t)=\mbox{sgn}(q(0))\biggl(2\sqrt{E}t+
  \frac{q(0)p(0)}{\sqrt{E}}\biggr)
  \label{43}
\end{equation}
which coincides with
the asymptotic form (\ref{41}), (\ref{42}), up to the permutation
 $1\leftrightarrow 2$. Finally we calculate $I_n$,
and $J_n$ explicitly
\begin{eqnarray}
  &&I_1=\Pi={p^\prime}_1+{p^\prime}_2,\qquad 
  I_2=\frac{\Pi^2}{4}-p^2-\frac{g}{q^2}={p^\prime}_1 {p^\prime}_2
  \nonumber\\
  &&J_1=2X={q^\prime}_1+{q^\prime}_2, \qquad
  J_2=X\Pi-qp={q^\prime}_1 {p^\prime}_2+{q^\prime}_2 {p^\prime}_1.
  \label{44}
\end{eqnarray}
Notice, however, that
\begin{equation}
  I_{2\,0}=q_1 q_2=X^2-\frac{q^2}{4}=
  {X^\prime}^2-\frac{{q^\prime}^2}{4}-\frac{g}{{p^\prime}^2}=
  {q^\prime}_1 {q^\prime}_2-\frac{g}{{p^\prime}^2}\neq I_{2\,0}(g(0)).
  \label{45} 
\end{equation}
In order to explain the appearance of the additional contribution it
is sufficient to calculate the asymptotic form of $q^2(t)$. {}From
eq.~(\ref{41}) one finds that there is an extra $O(1)$ piece
 coming from
the product of the first and the third terms on the right-hand side of
eq.~(\ref{41}). It coincides exactly with what is needed.

To conclude this section we show that, in the general case, the canonical
transformation (\ref{12}), which we denote by $C(\omega)$, does not 
depend on $\omega$.  Notice that 
$ C(\omega)\circ C^{-1}(\omega^\prime)$
leaves $I_n(g=0)$ and $J_n(g=0)$ invariant. Therefore it is simply
equal to a permutation of particles and, depending continuously on
$g$, must be an identity.

\section{The quantum case} \label{sec3}
In this section we construct a unitary operator which transforms the
$N$-particle 
quantum Calogero--Moser system into a system of free particles. 
Let $q_i$, $p_i$ be the canonical Heisenberg operators, 
$[q_i,p_i]=i\hbar \delta_{i\,j}$, $i,j = 1,2,\ldots, N$. 
Then the operators $T_0, T_\pm$
($\widetilde{T}_0, \widetilde{T}_\pm$ resp.), given by
\begin{eqnarray}
  &&T_+=\frac{1}{\omega\hbar}H_{CM}, \qquad
  \widetilde{T}_+=\frac{1}{\omega\hbar} H_{CM}(g=0) \nonumber\\
  &&T_-=\frac{\omega}{2\hbar} \sum_{i=1}^Nq_{i}^{2}=\widetilde{T}_-
  \label{47} \\
  &&T_0=\frac{1}{4\hbar} \sum_{i=1}^N (q_i p_i + p_i q_i)=\widetilde{T}_0,
  \nonumber
\end{eqnarray}
where $H_{CM}$ is the quantum Calogero-Moser Hamiltonian (\ref{1}), 
generate the $su(1,1)$ algebra, i.e.  $[T_0 ,T_\pm]=\pm iT_\pm$,
$[T_-,T_+] = 2iT_0$. 
\note{\begin{eqnarray}
  &&T_\pm=-T_1\pm iT_2 \nonumber\\
  &&T_0=-iT_3 \label{48}
\end{eqnarray}
Then 
\begin{equation}
  [T_i,T_j]=i\varepsilon_{i\,j\,k}T_k \label{49}
\end{equation}}
The unitary transformation corresponding to the canonical 
transformation (\ref{12})
is provided by the operator 
\begin{eqnarray}
  &&U=\mbox{e}^{-i\pi\widetilde{T}_1}\mbox{e}^{i\pi T_1}, \label{50}
\end{eqnarray}
where $T_1 = -\frac{1}{2}(T_++T_-)$ and $\widetilde{T}_1 = -\frac{1}{2}
(\widetilde{T}_++\widetilde{T}_-)$. $U$ is a well-defined unitary 
operator, which  describes the time evolution
$(t=\frac{\pi}{2\omega})$ generated by the self adjoint Hamiltonian of
the Calogero system followed by the time evolution
$(t=-\frac{\pi}{2\omega})$ of the set of $N$ harmonic oscillators. 
Actually, $U$ is
simply the time evolution operator for the Calogero model in the
interaction picture, calculated for $t=\frac{\pi}{2\omega}$.

Now we prove that $U$ has the same properties as its classical
counterpart. First  note that:
\begin{description}
\item{(i)} $I_n$, $J_n$ and $\phi_k$ can be defined by
``naive'' quantization of the relevant classical formulae; in fact,
there is no ordering problem here;
\item{(ii)} the transformation properties of $I_n$ and $J_n$ can be
  derived along the same lines as in Section 2 once the formula
  (\ref{19}) is shown to hold also in the quantum case.
\end{description}
To show that eq.\ (\ref{19}) is valid in the quantum case too, 
we use the normal ordering
technique of Ref.\ \cite{gon??}. For any (analytic) function $f(q,p)$
of canonical variables $q_i,p_i$, we define the operator $:\!f(q,p)\!:$,
which is obtained by replacing the canonical variables in the power
series expansion of $f$ by the relevant Heisenberg 
operators with all the 
$q_i$ preceding the $p_i$. The product of two such normally ordered
operators can be expressed as a normally
ordered operator via \cite{gon??}:
\begin{equation}
  :\!f(q,p)\!::\!g(q,p)\!:
  =:\!f\left(q,p-i\hbar\frac{\partial}{\partial r}\right)
  g(r,p)\Biggl\lvert_{r=q}\hspace{-4mm}\!:=:\!g\left(q-i\hbar \frac{\partial}{\partial r},p\right)
  f(q,r)\Biggl\lvert_{r=p}\hspace{-4mm}\!:. \label{51}
\end{equation}

Consider the Heisenberg equations for $:\!\phi_k(q,p)\!:$ with 
$\phi_k(q,p)$ given by (\ref{20})
\begin{eqnarray}
  i\hbar\frac{d:\!\phi_k(q,p)\!:}{dt}\!\!\!& = &\!\!\![:\!\phi_k(q,p)\!:,:\!H_C\!:]\nonumber\\
&=&\!\!\!\!
  :\!H_C\left(q-i\hbar\frac{\partial}{\partial r},p\right)
  \phi_k(q,r)\Biggl\lvert_{r=p}\hspace{-4mm}: 
  -:\!H_C\left(q,p-i\hbar\frac{\partial}{\partial r}\right) 
  \phi_k(r,p)\Biggl\lvert_{r=p}\hspace{-4mm}: \label{52}
\end{eqnarray}
The right-hand side of (\ref{52}) reads
\begin{eqnarray}
  &&\hspace{-1.5cm}
  :\!\Biggl(\sum\limits_{i}\frac{p_{i}^{2}}{2}+
  \frac{g}{2}\sum\limits_{i\neq j}
  \frac{1}{\left(\left(q_i-q_j\right)-
  i\hbar\left(\frac{\partial}{\partial p_i}-
  \frac{\partial}{\partial p_j}\right)\right)^2}+\frac{\omega^2}{2}
  \sum\limits_{i} 
  \biggl(q_i-i\hbar\frac{\partial}{\partial p_i}\biggr)^2\Biggr)
  \phi_k(q,p)\!:\nonumber\\
  &&\hspace{-1.5cm}
  -:\!\Biggl(\sum\limits_{i}\frac{\left(p_{i}-
  i\hbar\frac{\partial}{\partial q_i}\right)^2}{2}+
  \frac{g}{2}\sum\limits_{i\neq j} \frac{1}{(q_i-q_j)^2}+
  \frac{\omega^2}{2}\sum\limits_{i}q_{i}^{2}\Biggr)\phi_k(q,p)\!: 
\label{53}
\end{eqnarray}
Expand (\ref{53}) in the powers of $\hbar$. The first nonvanishing term 
comes out as
\begin{equation}
  i\hbar :\!\bigg\{\phi_k(q,p),H_C\biggr\}\!:, \label{53a}
\end{equation}
so that the time evolution of $:\!\phi(q,p)\!:$ is given by  the
normally-ordered formula (\ref{19}), provided that all the other terms
in the expansion of $(\ref{53})$
vanish. To show that this is indeed the case, first
note that the expansion terminates on $\hbar^2$ terms. Indeed, for $n>2$, 
$\left(\frac{\partial}{\partial p_i}-\frac{\partial}{\partial p_j}\right)^n$ 
necessarily contains terms of the form 
$\frac{\partial^{k+l}}{\partial p_{i}^{k} \partial p_{j}^{l}}$ with 
$k\geq 2$ or $l\geq 2$ while $\phi_k(q,p)$ is multilinear in $p$. The
order $\hbar^2$ contribution to (\ref{52}) reads
$$
  \hbar^2 :\!3g\ij \frac{1}{\left(q_i-q_j\right)^4}
  \frac{\partial^2 \phi_k(q,p)}{\partial p_i \partial p_j}+
  \frac{1}{2}\sum_{i=1}^N
  \frac{\partial^2 \phi_k(q,p)}{\partial q_{i}^{2}}\!:. 
$$
Using the explicit form of $\phi_k(q,p)$, one checks easily that
both terms cancel each other.

Finally, we show that $U$ does not depend on $\omega$. Arguing in
a similar way as in Section 2 we conclude that 
\begin{equation}
  U(\omega)U^{-1}(\omega^\prime)=
  \mbox{e}^{i\sigma(\omega,\omega^\prime)} I. \label{55}
\end{equation}
Taking the adjoint of eq.\ (\ref{55}) one obtains
\begin{equation}
  \sigma(\omega,\omega^\prime)=-\sigma(\omega^\prime,\omega). \label{56}
\end{equation}
On the other hand the identity
\begin{equation}
  \left(U(\omega) U^{-1}(\omega^{\prime\prime})\right)
  \left(U(\omega^{\prime\prime}) U^{-1}(\omega^\prime)\right)=
  \mbox{e}^{i\sigma(\omega,\omega^\prime)} I \label{57}
\end{equation}
together with the continuity of $\sigma$ imply 
\begin{equation}
  \sigma(\omega,\omega^{\prime\prime})+
  \sigma(\omega^{\prime\prime},\omega^\prime)=\sigma(\omega,\omega^\prime).
  \label{58}
\end{equation}
The general solution to (\ref{56}) and (\ref{58}) reads
\begin{equation}
  \sigma(\omega,\omega^\prime)=\rho(\omega)-\rho(\omega^\prime). \label{59}
\end{equation}
Apart from $\omega$, the function  $\rho$ can depend on $g$ and
$\hbar$ only. However, $\rho$ is dimensionless, therefore the 
dimensional analysis yields  that
$\rho$ cannot depend on
$\omega$. Therefore $\sigma(\omega,\omega^\prime)\equiv 0$ and
$U(\omega)$ is $\omega$-independent.

We complete the paper by illustrating the above discussion by the
reference to the two-particle quantum Calogero--Moser model. 
Passing to the relative
coordinates and ignoring the centre-of-mass motion one easily finds
the normalised eigenfunctions of the Calogero Hamiltonian 
$H_{C} = p^2 + \frac{g}{q^2}
+\frac{\omega^2}{4}q^2$, and the corresponding 
energy levels (for definiteness, we consider the
fermionic case)
\begin{eqnarray}
  &&\phi_{n}^{(a)}=\mbox{sgn}(q)\sqrt{\frac{n!}{\Gamma (a+n+1)}} 
  \biggl(\frac{\omega}{2\hbar}\biggr)^{\frac{a+1}{2}} 
  \left|q\right|^{a+\frac{1}{2}} 
  L_{n}^{a}\biggl(\frac{\omega q^2}{2\hbar}\biggr) 
  \mbox{e}^{-\frac{\omega q^2}{4\hbar}}, \nonumber\\
  &&E_n=\hbar\omega (2n+a+1), \label{60}
\end{eqnarray}
where $a\equiv \frac{1}{2}\sqrt{1+\frac{4g}{\hbar^2}}$, and $L^a_n$ are
the Laguerre polynomials. The propagator
\begin{equation}
  K^{(a)}(q,q^\prime,t)=
  \sum_{n=0}^\infty\phi_{n}^{(a)}(q)\overline{\phi_{n}^{(a)}(q^\prime)}
  \mbox{e}^{-\frac{iE_n t}{\hbar}}
\end{equation}
can be calculated using the well-known properties of the Laguerre
polynomials \cite[pp.\ 1037--1039]{gra71}
\begin{equation}
  K^{(a)}(q,q^\prime,t)=\frac{1}{2}
 \mbox{sgn}(qq^\prime)\bigl(\frac{\omega}{2\hbar}
e^{-\frac{\pi}{2}(a+1)i}\bigr) 
  \left|qq^\prime\right|^{\frac{1}{2}}
  \frac{\exp\{\frac{i\omega}{4\hbar}
  \left(q^2+{q^\prime}^2\right)\cot\omega t\}}{\sin\omega t} 
J_a\left(\frac{\omega}{2\hbar}
  \frac{|qq^\prime|}{\sin\omega t}\right) , 
  \label{62}
\end{equation}
where $J_a$ is the Bessel function of the first kind. 
The kernel of the unitary transformation (\ref{50}) can be rewritten as 
\begin{equation}
  U(q,q^\prime)=\int_{-\infty}^{\infty} dq''
  K^{(\frac{1}{2})}\left(q,q'',t=-\frac{\pi}{2\omega}\right) 
  K^{(a)}\left(q'',q',t=\frac{\pi}{2\omega}\right). \label{63}
\end{equation}
This can be developed further with the help of eq.\ (\ref{62}) and the 
explicit form of $J_{\frac{1}{2}}(x) = \sqrt{\frac{2}{\pi x}}\sin x$
\begin{equation}
  U(q,q')=\left(\frac{1+i}{4\sqrt{\pi}}\right)
  \mbox{e}^{-\frac{i\pi a}{2}} \left|q'\right|^{\frac{1}{2}}
  \mbox{sgn}(qq')\int_{-\infty}^{\infty}d\lambda 
  \left|\lambda\right|^{\frac{1}{2}} \sin(\left|\lambda q\right|)
  J_a(\left|\lambda q'\right|).
\end{equation}
In order to check the correctness of the above expression 
compute
\begin{equation}
  \int_{-\infty}^{\infty} dq' U^{-1}(q,q')
  \phi(q')=\int_{-\infty}^{\infty} dq' \overline{U(q',q)} \phi(q'),
  \label{65}  
\end{equation}
where $\phi(q)=\sin(kq)$ is a free fermionic wave function. A simple
computation gives 
\begin{equation}
  \int_{-\infty}^{\infty} dq' U^{-1}(q,q') \phi(q')=
  \frac{(1-i)}{2}\sqrt{\pi} \mbox{e}^{\frac{i\pi a}{2}}
  \mbox{sgn}(q) \left|kq\right|^{\frac{1}{2}} J_a(k\left|q\right|).
  \label{66} 
\end{equation}
The right-hand side is a fermionic solution to the Schr\"{o}dinger
equation for the Calogero--Moser model, corresponding to the energy 
$E=\hbar^2 k^2$.\vspace{1cm}

\noindent {\bf Acknowledgments}

\noindent We are grateful to P. Kosi\'nski for numerous fruitful 
discussions and
suggestions. S. Giller and M. Majewski are acknowledged for their
valuable remarks. The research of T.B. and C.G. is supported by the
British Council grant WAR/992/147. The research of P.M. is supported by
the KBN grant No. 2 PO3B 076 10.


\end{document}